\documentclass[aps,prl,preprint,groupedaddress]{revtex4}

\usepackage{amssymb}

\newcommand{\ZZ}{\mathbb{Z}}
\newcommand{\RR}{\mathbb{R}}
\newcommand{\CC}{\mathbb{C}}
\newcommand{\tr}{\mathop{\rm Tr}}
\newcommand{\im}{\mathop{\rm Im}}
\newcommand{\re}{\mathop{\rm Re}}
\newcommand{\sphere}{\mathbb{S}}

\begin{document}


\title{Compactified
strings as quantum statistical partition
function on the Jacobian torus}


\author{Marco Matone}
\author{Paolo Pasti}
\author{Sergey Shadchin}
\author{Roberto Volpato}
\affiliation{Dipartimento di Fisica ``G. Galilei'' and Istituto
Nazionale di Fisica Nucleare Universit\`a di Padova,
Via Marzolo, 8 -- 35131 Padova, Italy}


\date{\today}

\begin{abstract}
We show that the solitonic contribution of toroidally compactified
strings corresponds to the quantum statistical partition function of
a free particle living on higher dimensional spaces. In the simplest
case of compactification on a circle, the Hamiltonian is the
Laplacian on the $2g$-dimensional Jacobian torus associated to the
genus $g$ Riemann surface corresponding to the string worldsheet.
$T$-duality leads to a symmetry of the partition function mixing
time and temperature. Such a classical/quantum correspondence and
$T$-duality shed some light on the well-known interplay between time
and temperature in QFT and classical statistical mechanics.
\end{abstract}


\maketitle


A puzzling feature of QFT concerns the crucial analogy with
classical statistical mechanics where the inverse temperature plays
the r\^ole of imaginary time. Time and temperature also mix by
performing analytic continuation along complex paths in the
path-integral. Another well-known analogy concerns the transition
amplitude for a particle for the time $it$ that coincides with the
classical partition function for a string of length $t$ at
$\beta=1/\hbar$. Even if such analogies follow as technical
properties, it is widely believed that they are deeply related to
the properties of space-time and should emerge in the string
context.

A step towards a better understanding of temperature/time and
classical/quantum dualities, would be finding a physical model where
these emerge naturally. In proposing such a model, we will start by
showing that string theory possesses a basic surprising
classical/quantum duality. It turns out that the {\it solitonic
sector} of toroidally compactified strings have a {\it dual} description as
{\it quantum statistical partition function} on higher dimensional
spaces, built in terms of the Jacobian torus of the string
worldsheet and of the compactified space. More precisely, in this
letter we will see that in the case of compactification on a circle
\begin{equation}\label{laprima}
\sum_{m,n\in\ZZ^g}e^{-\beta S_{m,n}}=\tr\, e^{-\beta
H}\ ,\end{equation} where $\beta=2R^2/\alpha'$, with $R$ the compactification
radius, and the Hamiltonian $H$ is $\Delta_{J_\Omega}/2\pi$, with
$\Delta_{J_\Omega}$ the Laplacian on the Jacobian torus $J_\Omega$
of the worldsheet. Eq.(\ref{laprima}) is just the direct consequence of
the stronger identity we will prove. Namely \begin{equation}\label{laprimab}H
\Psi_{m,n}=S_{m,n}\Psi_{m,n}\ ,\end{equation} that is the set $\{
S_{m,n}|m,n\in\ZZ^g\}$ {\it coincides} with the spectrum of $H$. As
will become clear from the construction, such a result seems to
capture a general property of string theories which is due to the
underlying geometrical properties of Riemann surfaces. In particular,
the correspondence considered in this letter admits natural generalizations to
compactifications on higher dimensional tori.

Remarkably, temperature/time duality naturally emerges as a
consequence of the complexified version of $T$-duality, a
fundamental feature of string theory 
\cite{GiveonFU}. By (\ref{laprima}) the standard
$T$-duality corresponds to the invariance, up to a multiplicative
term given by powers of $\beta$, of the partition under inversion of
the temperature \begin{equation}\label{zero} \beta\longrightarrow 1/\beta \ .\end{equation}
Complexification of $\beta$ has basic motivations which interplay
between physics and geometry. Let us first consider the time
translate of an observable $A$,
$\alpha_t(A)=e^{iHt} A e^{-iHt}$,
and the expectation value
$$
\omega_\beta(A)=\tr\, \rho_\beta A=Z_{stat}(\beta)^{-1}\tr\, A
e^{-\beta H} \ .
$$ By invariance of the
trace under cyclic permutations we get
$\omega_\beta((\alpha_t(A)B)=\omega_\beta(B e^{iH(t+i\beta)} A
e^{-iH(t+i\beta)})$, so that
$$
\omega_\beta((\alpha_t A)B)=\omega_\beta(B\alpha_{t+i\beta}(A))\ ,
$$
that is time evolution is invariant, upon commutation, under an imaginary shift of the
time which is inversely proportional to the temperature. In such a
context the complexification of $\beta$ naturally appears in
globally conformal invariant QFT \cite{NikolovYG}. It is worth noticing that both in
\cite{NikolovYG} and in the BC
system \cite{BC}
the KMS (Kubo-Martin-Schwinger) states \cite{Haag,CMuno,CMdue} play a crucial r\^ole. In
particular, in the limit of
$0$-temperature the KMS states may be used to define the concept of
point in noncommutative space.

A simple but important property of the temperature is its positive
definiteness. This becomes transparent once time and temperature are
naturally combined in a unique variable $\tau$ as suggested by the
above analogies, namely \begin{equation}\label{uno}\beta={1\over k_B
T}=\im \tau \ , \qquad {t\over \hbar}=-\re\tau \ .\end{equation}  As
we will discuss later, complexification of $\beta$ is intimately
related to the positivity of the temperature. Complexifying a
variable taking positive values has a geometrical motivation that
may lead to a geometrical understanding of the emergence of time. In
particular, as will be illustrated, positivity of the temperature
combined with the time variable to build the complex $\tau$, leads
to consider $\tau$ as the torus modular parameter.

An apparently unrelated topic concerns the Riemann mapping theorem,
which plays a central r\^ole in uniformization theory and in the
theory of univalent functions. According to such a theorem there is
a unique analytic function $w=f(z)$ mapping a simply connected
region of $\CC$ one-to-one onto the disk $|w|<1$ such that
$f(z_0)=0$ and $f'(z_0)>0$. As a consequence any two simply
connected regions except $\CC$ itself, can be mapped conformally
onto each other. While $\CC$ is the universal covering of the torus,
any simply connected domain of $\CC$ can be mapped by a locally
univalent function to the upper half-plane, the universal covering
of negatively curved Riemann surfaces. Positivity of the temperature
and its combination with time to build $\tau$ leads to the geometry
of the upper half-plane that, unlike the case of the Riemann sphere,
is characterized by the identification of its automorphism group,
$PSL(2,\RR)$, with the isometry group of its natural metric, the
Poincar\'e metric. Such a group acts by linear fractional
transformations
$
\tau\rightarrow (A\tau+B)/(C\tau+D)
$
that, according to (\ref{uno}), correspond to mixing the r\^ole of
temperature and time.

Similar transformations appear in several physical models, for
example in Seiberg-Witten theory. Again, positivity, that this time
is due to the coupling constant, plays a crucial r\^ole.

Geometrically $\tau$ can be seen as the modular parameter of a
torus, and then related to a flat geometry. On the other hand, it
can be seen as the inverse of the uniformizing map from a negatively
curved Riemann surface to the upper half-plane. In this way to each
point on a negatively curved Riemann surface one may associate an
elliptic curve. It should be stressed that negatively curved Riemann
surfaces are much more reach in nature and it may be convenient to
consider punctured versions of the torus rather than the torus
itself. Such a dual r\^ole for $\tau$, connecting flat and
negatively curved Riemann surfaces, is particularly transparent once
one considers the $\zeta$-function
$$
\zeta(s)=\sum_{m,n\in \ZZ}\frac{(\im\tau)^s}{|m+n\tau|^{2s}}
\ .
$$
The elements of the set $\{|m+n\tau|^{2}/\im\tau\mid m,n\in\ZZ\}$
are the eigenvalues of the Laplacian on the torus, so that
$e^{-\zeta(0)'}$ defines the determinant of the Laplacian on the
torus. On the other hand, $(\im\tau)^s$ can be seen as eigenfunction
of the Poincar\'e Laplacian. The summation then can be seen as a sum
on all $PSL(2,\ZZ)$ transformations of $({\rm Im}\,\tau)^s$
guaranteeing the invariance under the uniformizing group
$SL(2,\ZZ)$. This dual nature of $\tau$ is at the heart of basic
mathematical structures also involving number theory, and appears in
several topics of physical interest, not only in string theory.

Let $\Sigma$ be a genus $g$ Riemann surface with a fixed basis
$\{\alpha_1,\ldots,\alpha_g,\beta_1,\ldots,\beta_g\}$ of the first
homology group $H_1(\Sigma,\ZZ)$ with intersection matrix
$\alpha_i\cdot\beta_j=\delta_{ij}$,
$\alpha_i\cdot\alpha_j=\beta_i\cdot\beta_j=0$. A dual basis
$\{\omega_1,\ldots,\omega_g\}$, of holomorphic $1$-differentials on
$\Sigma$ can be chosen with normalization
$\int_{\alpha_i}\omega_j=\delta_{ij}$, $i,j=1,\ldots,g$, whereas the Riemann period matrix
$\Omega_{ij}=\int_{\beta_i}\omega_j$ can be proved to be symmetric with $\im\Omega>0$, and
depends on the complex structure of $\Sigma$. Conversely, the
Riemann period matrix completely determines the complex structure of
the corresponding Riemann surface (although for $g\ge 4$ not every
symmetric matrix with positive-definite imaginary part is the
period matrix of a Riemann surface).

Let us consider a scalar field theory on $\Sigma$ with the
one-dimensional target space compactified to a circle
$\sphere^1=\RR/2\pi R\ZZ$ of radius $R$. The partition function is
defined as a path-integral
$$Z(\beta)=\int_{(\Sigma,\sphere^1)} dX e^{-\beta S}\ ,$$
where $(\Sigma,\sphere^1)$ is the space of
maps from $\Sigma$ to $\sphere^1$ and \begin{equation}\label{stringact}S[X]=\frac{1}{4\pi R^2}\int_\Sigma
\partial X\bar{\partial} X\ ,\end{equation} with the normalization chosen for later reference. By setting
$\beta=2R^2/\alpha'$, this path-integral corresponds to the $g$-loop
contribution of a string theory with target space $\sphere^1$. Each
function $X:\Sigma\to\sphere^1$ satisfies the condition
$$X(z+p^t\alpha+q^t\beta)=X(z)+2\pi R(m^t p-n^t q)\ ,$$
$p,q\in \ZZ^g$,
where the winding numbers $m,n\in\ZZ^g$ label the different
solitonic sectors of $(\Sigma,\sphere^1)$. Let us split $X$ into
classical and quantum parts
$X=X^{cl}+X^{q}$, where $X^{cl}$ satisfies the classical equation of motion
$$\Delta X^{cl}=0\ ,$$
so that $X^{cl}$ is a harmonic function on $\Sigma$. It is worth noticing that there exists a unique
harmonic function for each solitonic sector, i.e. for each pair $(m,n)\in\ZZ^{2g}$. Therefore, the splitting can
be performed in such a way that the quantum contribution $X_q$ is a real
single-valued function on $\Sigma$, that is
$X_q(z+p^t\alpha+q^t\beta)=X_q(z)$. It follows that the path integral splits into a sum over
the inequivalent sectors labeled by solitonic numbers
$(m,n)\in\ZZ^{2g}$ times the functional integral over the space of
real single-valued functions on $\Sigma$
$$Z(\beta)=\sum_{m,n\in\ZZ^g}e^{-\beta S_{m,n}}\int_{(\Sigma,\RR)} dXe^{-\beta S}\ ,$$
whereas the mixed terms with classical times quantum part vanish.
The harmonic function in the sector $(m,n)\in\ZZ^{2g}$ is
$$X^{cl}_{m,n}(z,\bar z)=\frac{\pi R}{i}(m+\bar\Omega n)^t(\im\Omega)^{-1}\int^z\omega+c.c.\ ,$$
so that, by using the Riemann bilinear relations
$$\int_\Sigma\omega_i\wedge\bar\omega_j=-2i\im\Omega_{ij}\ ,$$ we obtain
$$S_{m,n}=S[X^{cl}_{m,n}]=\pi (m+\bar\Omega
n)^t(\im\Omega)^{-1}(m+\Omega n)\ .$$ Therefore, the partition
function is $Z=Z_{sol}Z_q$, where
\begin{eqnarray}\label{quant}&&Z_{sol}(\beta)=\sum_{m,n\in\ZZ^g}e^{-\beta\pi
(m+\bar\Omega n)^t(\im\Omega)^{-1}(m+\Omega
n)}\ ,\\
&&Z_q(\beta)=\left(\frac{A_\Sigma}{\det'(-\beta\Delta)}\right)^{1/2}\ ,\nonumber\end{eqnarray}
and $
A_\Sigma=\int_{\Sigma}\sqrt g$.

On the other hand, let us consider the quantum statistical partition
function at the temperature $T$ on a $g$-dimensional complex torus
$J_\Omega=\CC^g/(\ZZ^g+\Omega\ZZ^g)$, for some symmetric $g\times g$
matrix $\Omega$, with $\im\Omega>0$,
$$Z_{stat}(\beta)=\tr\, e^{-\beta H}\ ,$$
where $\beta=1/k_BT$, $H\equiv\Delta_{J_\Omega}/2\pi$ and
$$\Delta_{J_\Omega}=-2\im\Omega_{ij}\frac{\partial}{\partial z_i}\frac{\partial}{\partial \bar z_j}\ ,$$
is the Laplacian on $J_\Omega$ with respect to the natural metric
$ds^2=(2\im\Omega)^{-1}_{ij}dz^id\bar z^j$. Here and in the
following, $\beta$ and $H$ are rescaled by some fixed length $L$ and
thought of as dimensionless quantities. A complete orthogonal basis
of eigenfunctions for $H$ is $\{\Psi_{m,n}\}_{m,n\in\ZZ^g}$, with
$$\Psi_{m,n}(z,\bar z)=e^{\pi(m+\bar\Omega n)^t(\im\Omega)^{-1}z-c.c.}\ .$$
Indeed, a trivial computation shows that
$$H\Psi_{m,n}=\lambda_{m,n}\Psi_{m,n}\ ,$$
with eigenvalues
$$\lambda_{m,n}=\pi(m+\bar\Omega n)^t(\im\Omega)^{-1}(m+\Omega n)\ ,$$
so that $S_{m,n}=\lambda_{m,n}$ and
\begin{equation}\label{stat}Z_{stat}(\beta)=\sum_{m,n\in\ZZ^g}e^{-\beta\pi(m+\bar\Omega
n)^t(\im\Omega)^{-1}(m+\Omega n)}\ .\end{equation}
Comparing Eqs.(\ref{quant}) and (\ref{stat}), we obtain the remarkable identity
Eq.(\ref{laprima}) between the classical contribution to the partition
function in a $2$-dimensional field theory and the quantum
statistical partition function of a free particle in a
$2g$-dimensional space. Note that the space in the statistical
theory is exactly the Jacobian torus of the Riemann surface where
the first theory is defined. In particular, in the case of genus
$1$, $\Sigma$ coincides with its Jacobian and we obtain a duality on
the same space.

By applying the Poisson summation formula
$$\sum_{m\in\ZZ^d}e^{-\pi (m+a)^tA(m+a)+2\pi im^tb}=(\det A)^{-\frac{1}{2}}\sum_{m\in\ZZ^d}e^{-\pi
(m+b)^tA^{-1}(m+b)-2\pi i(m+b)^ta}\ ,$$
to the sum over $m$ in
$Z_{sol}(\beta)$, one obtains
$$Z_{sol}(\beta)
=\Big(\frac{\det\im\Omega}{\beta^g}\Big)^{1/2}\sum_{m,n\in\ZZ^g}e^{\pi[-\beta^{-1}
m^t(\im\Omega) m-\beta n^t(\im\Omega)n+2 im^t(\re\Omega) n]}=
\beta^{-g}Z_{sol}(1/\beta)\ ,$$
 where the correspondence $\beta\to
1/\beta$ implies $R\to R'=\alpha'/2R$, which is the standard
$T$-duality for string theory on a circle. In the case of $Z_{stat}(\beta)$, the same calculation leads to
$$Z_{stat}(\beta)=\beta^{-g}Z_{stat}(1/\beta)\ ,$$
which has to be interpreted as a hot-cold duality $T\to
T'=T^2_{sd}/T$, where $T_{sd}$ is the self-dual temperature $T_{sd}=\hbar c/k_BL$. Note that, by setting
$L\sim(\alpha'/2)^{1/2}$, a correspondence can be established between the energy scales in the string
theory model and in the quantum statistical one, relating the fixed points (self-dual radius and
temperature, respectively) under the duality $\beta\to 1/\beta$.

>From the point of view of
statistical mechanics, it is natural to consider the complexification
\begin{equation}\label{taubeta}\tau=-t+i\beta\ ,\end{equation} where $t$ denotes the time,
so that we can define
$$Z_{stat}(\tau)=\tr\,e^{i\tau H}\ .$$
By Poisson summation, we obtain the duality
$$Z_{stat}(\tau)= (-i\tau)^{-g} Z_{stat}(-1/\tau)\ ,$$
that mixes time and temperature, namely
$$
t\rightarrow \frac{-L^2}{c^2[t^2+(\hbar/k_BT)^2]}t\ , \quad
\frac{1}{T}\rightarrow\frac{L^2}{c^2[t^2+(\hbar/k_BT)^2]}\frac{1}{T} \ ,
$$
where the dimensional parameters have been recovered.

We have seen that the classical sector of the compactified string on
$\sphere^1$ has a quantum statistical description on the Jacobian
torus. This corresponds, like the original string, to a first
quantized theory. Also note that restricting the eigenfunctions of
$H$ to the image of $\Sigma$ in $J_\Omega$ provides a direct link
between $\Psi_{m,n}$ and $X^{cl}_{m,n}$, namely
\begin{equation}\label{vertex}\Psi_{m,n}(\textstyle{\int^z\omega,\overline{\int^z\omega}})=e^{\frac{i}{R}X^{cl}_{m,n}}
\ ,\end{equation} resembling a classical vertex. This may
indicate that the dual description of the classical contribution is
the facet of a more general dual description of the full string. For
example, the fact that the Laplacian acting on sections of the torus
and the one acting on sections of its Jacobian coincide, implies a
functional relation between the classical and quantum sectors of the
string (coming from the heat equation). This suggests that even the
quantum sector of the string admits a dual description which may
extend to higher genus as well. Understanding such an extension
means to investigate the possible relations, for any $g$, between
the spectrum of $\Delta_{J_\Omega}$ and the one of $\Delta$; a
problem that, as we will comment below, is of considerable
mathematical interest. For the time being we note that, due to the
appearance of $({\rm Im}\,\Omega)^{-1}_{ij}$, a distinguished r\^ole
may be played by the Bergman metric
$$
ds^2=\frac{1}{g}\sum_{i,j=1}^g\omega_i(z)({\rm
Im}\,\Omega)^{-1}_{ij}\overline{\omega_j(z)} \ .
$$
We note that the equality (\ref{laprima}) between the sum over the
topologically non-trivial states of a 2-dimensional sigma model and
the trace of operators over the Hilbert space of a free particle on
the Jacobian torus, is reminiscent of electric-magnetic duality in
$\mathcal{N}=2$ SYM theory. In the latter, the solitonic objects are
the 't Hooft-Polyakov monopoles and Julia-Zee dyons, and the
elementary objects are gluons. In our model Eq.(\ref{taubeta}) can be
thought of as the complex coupling constant. Its analog in SYM
theory is $\tau = \frac{\Theta}{2\pi}+\frac{4\pi i}{g^2}$. However, while electric-magnetic duality is between weakly coupled
and strongly coupled regimes of two different theories (or of the
same theory, if the theory is self-dual), in our model the duality
is established between the same regimes of different theories. The
way out is to note that by the composition of Eq.(\ref{laprima}) and
$T$-duality (\ref{zero}), precisely reproduces what we expect for an analog
of the electric-magnetic duality.

Let us further comment the complex combination of time and temperature. For the quantum
theory to be unitary the action must be real. The imaginary part of
the action leads to non-conservation of probability, as it is clear
from the form of the Feynman weight $e^{\frac{i}{\hbar}S}$. The
probability grows with the entropy, so the imaginary part of the
action should be negative. Moreover, its presence leads to
decreasing of the information, and therefore to increasing of the
entropy. A possible interpretation is to consider the imaginary part of the
action as the entropy: $S = \mathcal{A} - i \mathcal{E}$. If we write
the action $\mathcal{A}$ in the momentum space as a function of momentum
and energy $E$ (which can be achieved after taking its Legendre
transform) and use $\partial \mathcal{A}/\partial E = t$, we
obtain
$$
\partial S/\partial E = t - i\beta = -\tau\ .
$$
This suggests that the combination of time and temperature at hands is the variable conjugated to
energy in the case when the action is complex.

Let us conclude by observing that an intriguing outcome is that in
the case of $g=1$, the Jacobian corresponds to the torus itself. In
this way the quantum and solitonic contributions both are expressed
in terms of the same Laplacian. An old problem in the theory of
Riemann surfaces is to find an analytic expression for the
determinant of Laplacian acting on degree zero bundles in which the
dependence on the period Riemann matrix appears explicitly. In
\cite{MatoneTJ} it was conjectured a relation between the determinant of the
Laplacian on the Jacobian and the one on the Riemann surface. This
would provide a functional relation between $Z_{stat}(\beta)$ and
the quantum contribution to the string partition function also in
the higher genus case, generalizing the relation for $g=1$. The
eigenvalues $\lambda_{m,n}$ also appear in considering the Laplacian
with respect to degenerate metrics \cite{MatoneTJ}, for which ramified
covering of the torus play a crucial r\^ole \cite{MatoneUY}. Ramified covering of the torus
correspond to a particular kind of
CM (complex multiplication) satisfied by the Riemann period matrix
\cite{MatoneUY}. Remarkably, such special Riemann surfaces also appear in
the null compactification of type-IIA string perturbation theory at
finite temperature \cite{GrignaniZM}. It is worth noticing that CM, which is a lattice condition, also
appears in the study of sigma models on Calabi-Yau manifolds
\cite{GukovNW}.

We note that the dimensionality of our model suggests an intriguing
relation with the string theory compactified on a Riemann surface of
unitary volume in string units, where the effective degrees of
freedom are
still the ones of a $2g$-dimensional theory 
\cite{SilversteinQF}. Presumably this is connected to the relation between Fuchsian
groups and Liouville theory and to the fact that, in particular
regimes, the commutators between the Fuchsian generators may be
negligible so leading to a homological description.

\begin{acknowledgments}
Work partially supported by the
European Community's Human Potential Programme under contract
MRTN-CT-2004-005104 ``Constituents, Fundamental Forces and
Symmetries of the Universe".
\end{acknowledgments}

\bibliography{newprl}

\end{document}